\documentclass[twocolumn,a4paper,final,technote]{IEEEtran}
\usepackage{amsmath,amssymb}
\usepackage{graphicx}
\usepackage{color}
\usepackage{flushend}
\newcommand{\er}{\epsilon_r}

\newcommand{\Vr}{\vec{r}}
\newcommand{\vrp}{\vec{r}^{\prime}}

\begin{document}
\title{Spatial bandlimitedness of scattered electromagnetic fields
}
\author{Uday K Khankhoje and Kushal Shah%
\thanks{U.~K.~Khankhoje and K.~Shah are with the Department of Electrical Engineering, Indian Institute of Technology Delhi, Hauz Khas, New Delhi 110016, India. E-mails: uday@alumni.caltech.edu and kkshah@ee.iitd.ac.in, respectively.}}%

\maketitle

\begin{abstract}
In this tutorial paper, we consider the problem of electromagnetic scattering by a bounded two-dimensional dielectric object, and discuss certain interesting properties of the scattered field. Using the electric field integral equation, along with the techniques of Fourier theory and the properties of Bessel functions, we show analytically and numerically, that in the case of transverse electric polarization, the scattered fields are spatially bandlimited. Further, we derive an upper bound on the number of incidence angles that are useful as constraints in an inverse problem setting (determining permittivity given measurements of the scattered field). We also show that the above results are independent of the dielectric properties of the scattering object and depend only on geometry. Though these results have previously been derived in the literature using the framework of functional analysis, our approach is conceptually far easier. Implications of these results on the inverse problem are also discussed.
\end{abstract}


\section{Introduction}

Scattering is a phenomenon where some form of a travelling wave excitation (light, sound, etc.) deviates from its original trajectory due to a change in the properties of the medium along its path \cite{ishimaru1991electromagnetic}. In the context of electromagnetic waves, given a field incident on an object of known permittivity, it is quite straightforward to calculate the scattered field in various directions. In cases where this cannot be done analytically, several computational methods can be employed. This is known as the forward scattering problem. The inverse problem, however, is more challenging. It consists of determining the unknown spatial permittivity of an object based on measurements of the scattered field \cite{colton2012inverse}.

In order to understand the properties of the scattered field, Bucci et al.~\cite{bucci1989degrees,bucci1997electromagnetic} considered the electric field integral equation (EFIE) approach, and noted that the integral operator in this case is compact. By invoking a theorem due to Kolmogorov and Fomine \cite{kolmogorov1973elements} concerning the properties of such an operator, it was deduced that the scattered field has a finite dimensional representation. Further, the singular values of the operator rapidly decay after a certain threshold, a property attributed to the analyticity of the operator \cite{hille1931characteristic}. Thus, it was concluded \cite{bucci1997electromagnetic} that the scattered field can be represented by a finite number of singular vectors, each associated with a singular value. This critical number was referred to as the \textit{degrees of freedom} of the scattered field, and in the two-dimensional case of a circular observation domain bounding a circular scatterer of radius $a$, this number was found to be equal to $2ka$, where $k$ is the free space wavevector \cite{bucci1989degrees}. Also see \cite{franceschetti2009capacity} for a lucid derivation of this decomposition.

The mathematical machinery used in the works of Bucci et. al is formidable, as it is rigorous. Instead, we present a much simpler route to the same results by invoking certain elegant properties of Bessel functions. We start in Section \ref{methods} by discretizing the EFIE, and derive expressions for the discrete Fourier transform of the field scattered by a bounded dielectric object in the case of transverse electric polarization. Using a key property of Bessel functions of fixed argument (demonstrated in Appendix \ref{Bessel}), namely that their amplitude monotonically decays to zero as the order is increased beyond a threshold value, we present our main results on the bandlimited nature of the scattered fields in Section \ref{results}. In this section, we also present numerical validation with the finite element method, and consider the case of an object illuminated by multiple incidence angles. We conclude in Section \ref{discussion} with a discussion on the hardware implications of our results, and make a few other connections.

\begin{figure}[htbp]
\begin{center}

\includegraphics[width=0.5\textwidth]{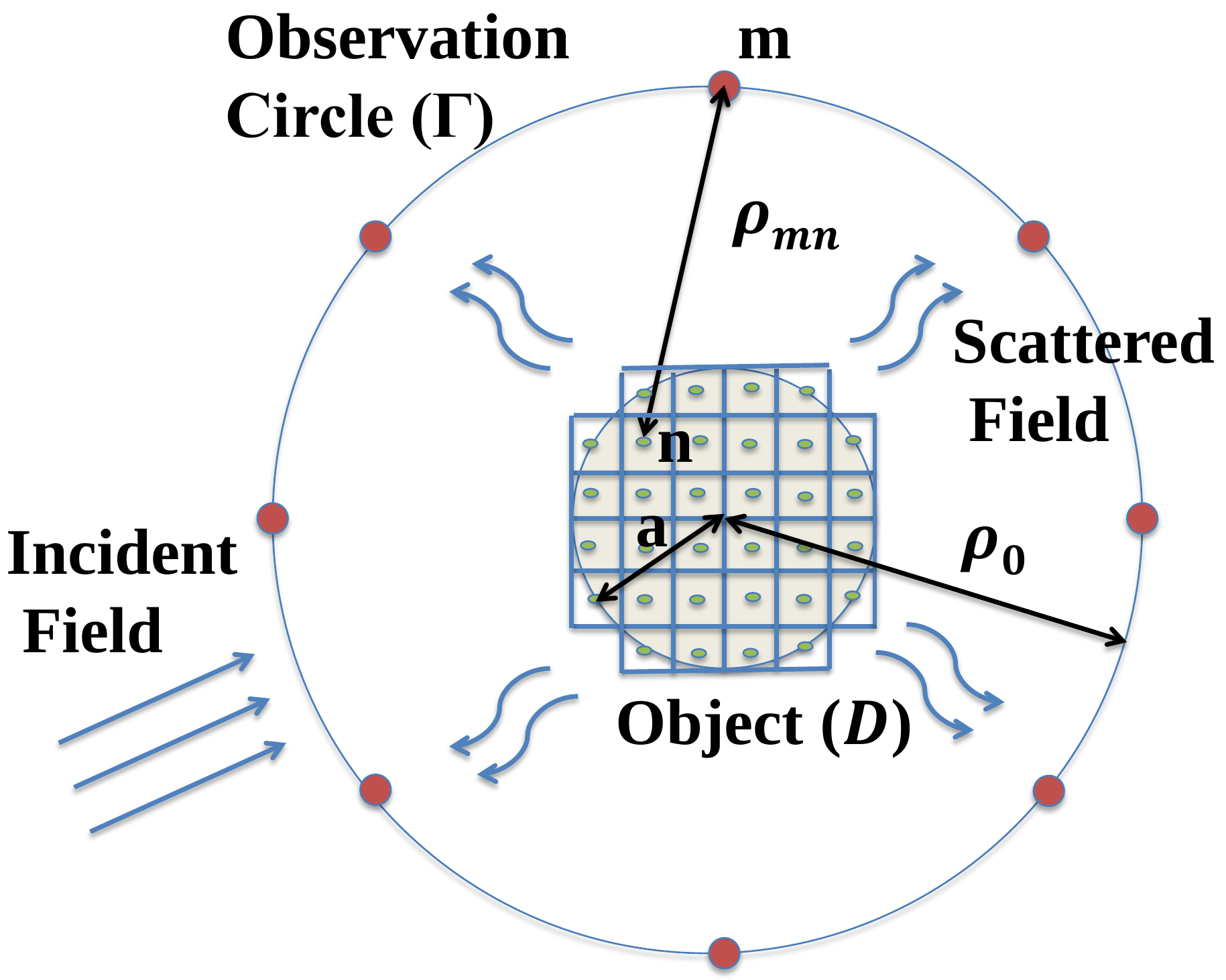}
\caption{A dielectric object $\mathcal{D}$ bounded within a circle of radius $a$ is illuminated by an incident field. The scattered field is measured on the circle $\Gamma$ of radius $\rho_0$. $n,m$ refer to discretized source and observation points, respectively.}
\label{fig1}

\end{center}
\end{figure}

\section{Methods}\label{methods}
\subsection{Deriving the scattered electromagnetic fields} 
\label{Forward-problem}

Consider a dielectric object bounded in a two-dimensional region of space, $\mathcal{D}$, whose relative permittivity, $\epsilon_r(\vec{r})$ is a function of space, immersed in a medium of constant (and real) relative permittivity, $\epsilon_b$. In this situation, if we consider a transverse electric (TE) polarization ($H_z=0$), the $z$-component of the electric field obeys the Helmholtz equation;
\begin{equation}
\nabla^2 E_z(\vec{r}) + k^2\epsilon_r(\Vr)E_z(\vec{r}) = 0
\end{equation}
where $k=2\pi/\lambda$ is the magnitude of the wavevector in free space, and $\epsilon_r(\Vr)$ in the relative permittivity at $\Vr$. The above equation can be recast as an integral equation in terms of the Green's function for the homogeneous medium (with uniform permittivity, $\epsilon_b$), $G(\Vr,\vrp)=(-j/4)H_0^{(2)}(k_b(\Vr-\vrp))$, where $H_0^{(2)}$ is the Hankel function of second kind and zero order, and $k_b =k\sqrt{\epsilon_b}$, as;
\begin{equation}
\label{intg_eqn}
E_z(\Vr) = E_z^i(\Vr) + k_b^2 \underset{\mathcal{D}}{\iint} G(\Vr-\vrp)\chi(\vrp)E_z(\vrp)\,d\vrp
\end{equation}
where $E_z^i$ is the incident field, and $\chi(\Vr)=\er(\Vr)/\epsilon_b-1$ is referred to as the dielectric contrast. The above EFIE is a Freedholm integral equation of the second kind.

Assume that the dielectric object, $\mathcal{D}$, is surrounded by a concentric observation circle, $\Gamma$, of radius $\rho_0$ shown in Figure~\ref{fig1}. To compute the scattered field ($E_z-E_z^i$) on $\Gamma$, we discretize Eq.~(\ref{intg_eqn}) by dividing the region $\mathcal{D}$ into $N$ equi-sized cells of uniform dielectric constant and $\Gamma$ into $M$ equispaced points. Following well known techniques for solving such equations \cite{richmod1965scattering,harrington1996field}, the scattered electric field ($v$) at the $m^{\text{th}}$ observation point on $\Gamma$ is related to the total field ($u$) at the N points of $\mathcal{D}$: 
\begin{align}
\label{scatf-2}
v_m &= \overset{N-1}{\underset{n=0}{\sum}}\, g_0 H_0^{(2)}(k_b\rho_{mn}) \chi_n u_n, \quad \text{where}\\
g_{0}&=-jk_b a_n\frac{\pi}{2} J_1(k_ba_n)
\end{align}
where $\rho_{mn}$ is the distance between the $m^{\text{th}}$ (observation) point on $\Gamma$ and $n^{\text{th}}$ (source) point in $\mathcal{D}$, $a_n$ is the radius of the circle with the same area as the $n^{\text{th}}$ cell of $\mathcal{D}$ \cite{richmod1965scattering} with dielectric contrast $\chi_n$, and $J_1$ is the Bessel function of the first kind and first order. Note that without loss of generality, $a_0=a_1=\dotsb=a_{N-1}$ and $g_0$ does not depend on $n$.

We now approximate the scattered field in a far-field setting ($\rho_0 >\lambda$). The cosine rule gives us that $\rho_{mn} =(r_n^2+\rho_0^2-2r_n \rho_0 \cos\theta_{mn})^{1/2}$, where $\theta_{mn}$ is the angle between the position vectors corresponding to the source and observation points, $\Vr_n$ and $\Vr_m$, respectively, and $|\Vr_m|=\rho_0$. Thus, in the far field, at least to first order, $\rho_{mn}$ can be approximated as $\rho_{mn}\approx \rho_0$ in the amplitude term and as $\rho_{mn} \approx \rho_0-r_n\cos\theta_{mn}$ in the phase term, following which (also using the large argument approximation of the Hankel function, $H_0^{(2)}(z) \approx \sqrt{2/(\pi z)}\exp{(-j(z-\pi/4))}$):
\begin{eqnarray}
\label{Hankel-approx}
H_0^{(2)}(k_b\rho_{mn}) & \approx & \left\{ \sqrt{\frac{2}{\pi k_b \rho_0}}\exp\left[j\left(\frac{\pi}{4}-k_b \rho_0\right)\right] \right\} \times \nonumber \\
&{}&\qquad \exp{(jk_b r_n\cos\theta_{mn})} \nonumber \\
&=& h_0 \exp{(jk_b r_n\cos\theta_{mn})}
\end{eqnarray} 
where $h_0$ is a constant independent of $m$ when the observation points are on a circle (the terms in the curly bracket above). Finally, the far field approximation for the scattered field can be written as;
\begin{equation}
\label{scat-ff}
v_m = g_0 h_0 \overset{N-1}{\underset{n=0}{\sum}}\, \chi_n u_n  \exp{(jk_b r_n\cos\theta_{mn})}
\end{equation}

\subsection{Discrete Fourier Transform of the Scattered Fields}
\label{DFT-plane}

We are now in a position to consider the $M$-point discrete Fourier transform (DFT), $\tilde{v}\,\in\mathbb{C}^M$, of the scattered electric field, $v$, as obtained in Eq.~(\ref{scat-ff}). The $k^{\text{th}}$ Fourier component is given by; 
\begin{eqnarray}
\label{dftscatf}
\tilde{v}_k &=& \overset{M-1}{\underset{m=0}{\sum}}\, \exp{\left(-j2\pi k\frac{m}{M}\right)}\times \nonumber \\
&{}&\qquad\overset{N-1}{\underset{n=0}{\sum}}\, g_0 h_0 \exp{(jk_b r_n\cos\theta_{mn})} \chi_n u_n .
\end{eqnarray}

A simple reordering of the order of summations in (\ref{dftscatf}) reveals that the inner summation is, in effect, the DFT of a plane wave sampled on a circle of radius $r_n$;
\begin{eqnarray}
\label{dftscatf_simple}
\tilde{v}_k & = & g_0 h_0 \overset{N-1}{\underset{n=0}{\sum}}\chi_n u_n \times \nonumber \\
&{}& \,\overset{M-1}{\underset{m=0}{\sum}}\, \exp\left(-j2\pi k\frac{m}{M}\right)\,\exp{\left(jk_b r_n\cos\theta_{mn}\right)} . 
\end{eqnarray}
This is so because the inner summation, in which the source position, $\Vr_n$, is fixed, contains $\theta_{mn}$ in the exponent. This $\theta_{mn}$ can be expanded as  $\theta_{mn}=(2\pi m/M-\theta_n)$, where $\theta_n$ is the angular position of the $n^{\text{th}}$ source point. Since the measurement position, $\Vr_m$, goes around a circle of radius $\rho_0$ as $m$ goes from $0$ to $(M-1)$, it is evident that $\theta_{mn}$ evenly samples points spanning $2\pi$.

\section{Results}\label{results}
\subsection{Bandwidth: Dependence on size}
\label{bandlimit-analytical}

The DFT of the scattered field as derived in Eq.~(\ref{dftscatf_simple}) can be simplified using the Jacobi-Anger expansion, as follows (see Appendix \ref{planewaveDFT} for details) ;
\begin{eqnarray}
\label{dftscatf_simple-2}
\tilde{v}_k & = & g_0 h_0 M \overset{N-1}{\underset{n=0}{\sum}}\, \chi_n u_n \overset{\infty}{\underset{q=-\infty}{\sum}}\,j^{k-qM} \times \nonumber \\
&{}& \quad  \exp{[-j (k-qM) \theta_{n}]}  J_{k-qM}(k_br_n).
\end{eqnarray}
A careful examination of the inner summand of Eq.~\eqref{dftscatf_simple-2} shows that for a fixed point in region $\mathcal{D}$ (i.e.~fixed $n$), the argument of the Bessel function is constant (equal to $k_b r_n$) and only its order (equal to $k-qM$) changes with $q$. 

The Bessel function, $J_k(x)$, has the property that $J_k(x) \sim 0$ for $|k| \ge \lceil{2|x|}\rceil$ (see Appendix \ref{Bessel}). Further, since the maximum possible value of $r_n$ is $a$ (for a dielectric bounded within a cylindrical region of radius $a$), it naturally follows that for $k\ge \lceil{2k_b a}\rceil$ the Fourier component, $\tilde{v}_k$ will be negligible. In other words, the DFT, $\tilde{v}$, is bandlimited to this value $\eta = \lceil{2k_b a}\rceil$. It must be mentioned that this bandlimit corresponds to an \emph{effective} bandwidth, as the Fourier coefficients for $\eta<k<M-\eta$ are negligible, but not identically zero \cite{bucci1997electromagnetic}.

We thus arrive at the same result as Bucci et.~al \cite{bucci1989degrees} regarding the degrees of freedom of the scattered field in terms of an effective bandwidth. We note that our approach of discretizing the EFIE is similar to one previously proposed \cite{chew1994inverse}, wherein, starting from a series representation of the Green's function, a truncated Fourier series for the scattered electric field is obtained; we essentially extend this idea and estimate the truncation number in terms of $\eta$.

\subsection{Bandwidth: Dependence on permittivity}

Since the internal field coefficients, $u_n$, and the contrast, $\chi_n$, are upper bounded in magnitude, the Fourier components in Eq.~(\ref{dftscatf_simple-2}) that are zero, will continue to be zero, regardless of the particular values of the field and contrast.

To see this  quantitatively, consider the $k^{\text{th}}$ Fourier coefficient of the scattered field from Eq.~(\ref{dftscatf_simple-2}). We consider the first half of the total $M$ coefficients; this is sufficient since the DFT is symmetric (see  Appendix \ref{planewaveDFT-symmetry}):
\begin{equation}
\label{dftscatf_simple-3}
\tilde{v}_k  =  g_0 h_0 Mj^{k} \overset{N-1}{\underset{n=0}{\sum}}\,\left(\chi_n\right) \left[ u_n \exp{(-j k\theta_{n})}J_{k}(k_br_n)\right] 
\end{equation}
Assuming the contrast to be bounded such that ${|\chi_n| \leq \kappa},\,\forall n$, and the field to be bounded such that ${|u_n| \leq \gamma},\,\forall n$, we apply the Cauchy-Schwartz inequality ($|\underset{n}{\sum}a_n\bar{b}_n|^2\leq \underset{n}{\sum}|a_n|^2 \underset{n}{\sum}|b_n|^2 $) to the above relation to obtain
\begin{align}
\nonumber
|\tilde{v}_k|^2 & \leq |g_0 h_0 M|^2 \overset{N-1}{\underset{n=0}{\sum}}\left| \chi_n \right|^2   \overset{N-1}{\underset{n=0}{\sum}} \left| u_n \exp{(-j k \theta_{n})}J_{k}(k_br_n)\right|^2 \\
&\leq |g_0 h_0 M|^2N(\kappa\gamma)^2 \overset{N-1}{\underset{n=0}{\sum}}\left|J_{k}(k_br_n)\right|^2
\end{align}

Thus, if the magnitude of the order of the Bessel function, $|k|$, is large enough such that $J_{k}(k_b a) \sim 0 $, we can see that $|\tilde{v}_k| \sim 0$ as long as $\kappa,\gamma$ are finite. In other words, the bandlimited nature of scattered field is independent of the object permittivity, and this bandlimit depends only on the size of the object relative to wavelength, $k_ba$.

\subsection{Numerical Results}

We use a two dimensional vector-element based finite element method (FEM) \cite{khankhoje2013} to compute the scattered electromagnetic fields in two different configurations. In both configurations, the scattering object is confined within a cylinder of radius $r=2\lambda$, the fields are computed on a radius of  $r=3\lambda$, the computational domain is terminated by applying a radiation boundary condition at a radius $r=4\lambda$, and the TE-polarized incident field makes an angle of $0^\circ$ with the $+x$ axis. For the numerical convergence of the FEM solution, the domain discretization must be on the order of $\lambda/20$, which results in a mesh having approximately 90,000 elements. In the first configuration, the cylinder has a uniform permittivity, $\epsilon_r = 4$, while in the second configuration, we allow the permittivity of the cylinder to be random such that $\epsilon_r$ for each element is a uniform random variable from 1 to 10. 

\begin{figure}[htbp]
\begin{center}

\includegraphics[width=0.5\textwidth]{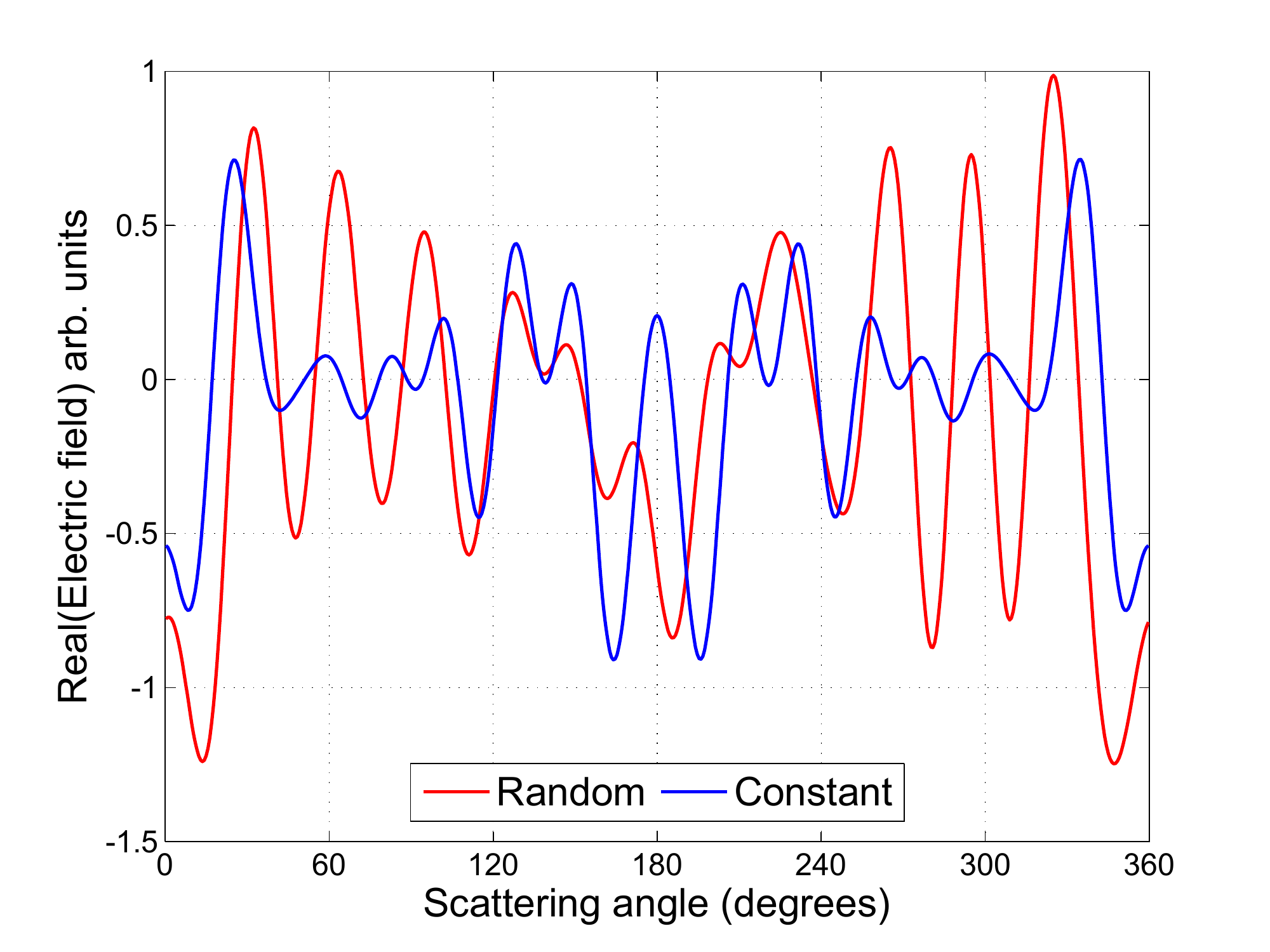}
\caption{FEM-computed real part of the scattered electric fields as a function of scattering angle as observed on a circle of radius $r=3\lambda$ for a constant permittivity ($\epsilon_r=4$) cylinder (blue curve) and for a cylinder with random permittivity (red curve). In the latter, the permittivity of each element in the cylinder is a uniform random variable such that $\epsilon_r \in [1,10]$ (see top-left inset of Figure \ref{fig-fft}).}
\label{fig-fields}

\end{center}
\end{figure}

\begin{figure}[htbp]
\begin{center}
\includegraphics[width=0.5\textwidth]{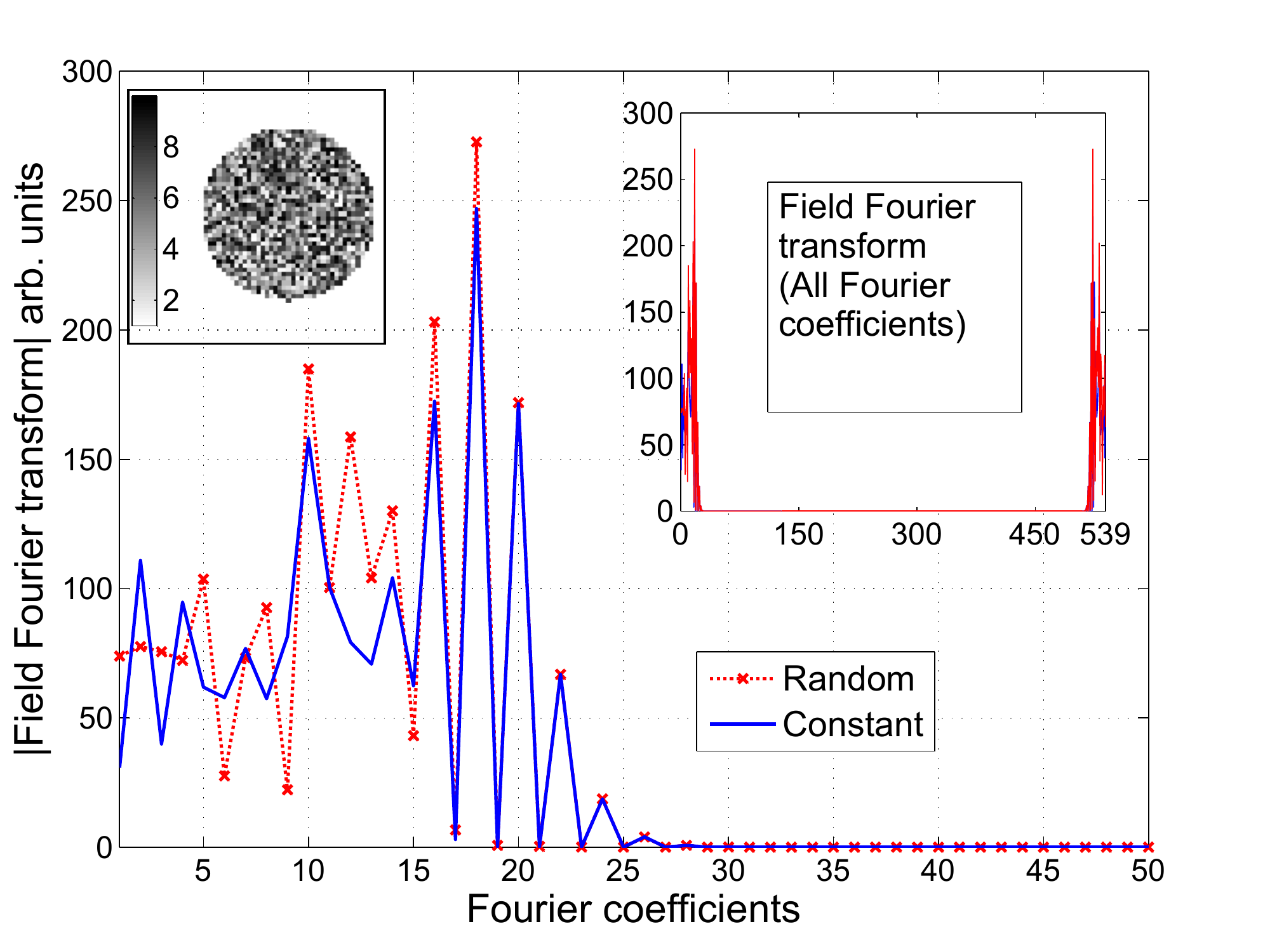}
\caption{Absolute value of the DFT of the scattered fields as shown in Figure \ref{fig-fields} for the two cylinders with constant (blue) and random (red) permittivity. The top-right inset shows all Fourier coefficients, clearly demonstrating the symmetric and bandlimited nature of the fields for both cylinders. The main graph shows the first 50 Fourier coefficients, and it is evident that beyond $k=26$ the Fourier coefficients have negligible magnitude. The inset on the top-left shows the permittivity of the random cylinder as a function of space.}
\label{fig-fft}
\end{center}
\end{figure}

The scattered fields in both cases are shown in Figure \ref{fig-fields}, and their corresponding spatial DFTs in Figure \ref{fig-fft}. We find that regardless of the constitutive permittivity of the scattering object, the DFT is bandlimited, and that the band limit matches very well with the analytical prediction of $\eta=\lceil 2k_b a \rceil$ from Section \ref{bandlimit-analytical}. Here, $a=2\lambda$ , which gives $\eta=26$ (see Figure \ref{fig-fft}). It is interesting to note that while the observation circle ($r=3\lambda$) is not in the far-field of the scattering object ($a=2\lambda$), the predicted cut-off matches very well with the FEM results (which do not assume any far-field approximation). This is because the idea of an effective bandwidth as derived by Bucci et al.~\cite{bucci1989degrees} does not require a far-field approximation, even though we assume it here to simplify the analysis.

\subsection{The case of multiple incidence angles}\label{multiple}
So far, we have considered the dielectric object to be illuminated by a single electromagnetic wave, and found that the scattered field can be represented in terms of $2k_ba$ Fourier components. In other words, in an inverse problem type setting, there will be $2k_b a$ linearly independent relations that can be used to relate the known value of the scattered field to the unknown permittivity of the object. In the language of the formalism presented in this paper, this amounts to determining $\chi_n,\,n\in[0,N-1]$ given the measurements $v_m,\,m\in[0,M-1]$ from Eq.~(\ref{scatf-2}).

A natural question that arises is: by what method can more information be gained about the object? It is conceivable that by introducing multiple incidence angles, further constraints can be imposed on the unknown object permittivity. However, we show that the number of such constraints can not be increased linearly with the number of incidence angles, and that there is an upper limit on the number of useful angles.

To see this, we go back to the integral equation formulation as in Eq.~(\ref{intg_eqn}) and note that this equation, when discretized, leads to a well-determined system of equations \cite{harrington1996field}, denoted by ${F\,u^i=b^i}$, where $F$ is a ``forward" matrix, $u^i$ consists of the internal field values for the $i^{\text{th}}$ incidence angle, and the incidence angle appears in the known vector, $b^i$. Explicitly, at the $k^{\text{th}}$ source point in the object domain, $b^i_k= \exp[-jk_b \rho_k\cos\theta_{ik}]$, where $\rho_k$ is the distance of the source from the origin and $\theta_{ik}$ is the angle between the source position vector, $\vec{r}_k$ and the incident wavevector, $\vec{k}_i$. We assume that the angular spacing between multiple incidence angles is uniform. The above matrix system has a solution, giving the field at the $n^{\text{th}}$ source point as;
\begin{equation}\label{fwd_inc}
u^i_n = \sum_{k=0}^{N-1} \left[F^{-1}\right]_{nk} \exp[-jk_b \rho_k\cos\theta_{ik}]
\end{equation}

Noting the similarities between Eq.~(\ref{scat-ff}) and Eq.~(\ref{fwd_inc}), it is at once clear that if the DFT of the above expression is taken w.r.t. the incident field index, $i$, the result would be a band limited expression, just as was shown with the DFT of Eq.~(\ref{scat-ff}) w.r.t. the observation point index, $m$ in Section \ref{DFT-plane}. Extending the analogy further, it is seen that this band limit is given by the expression;
\begin{equation}
\eta_p = \underset{\rho_k}{\text{max}} \lceil{2k_b \rho_k}\rceil = \lceil{2k_b a}\rceil.
\end{equation} 
Thus, only upto $\eta_p$ incidence angles are useful in imposing independent constraints on the unknown permittivity of the object. Beyond this number, the field can be reconstructed using $\eta_p$ sampled values and a suitable interpolation scheme \cite{candocia1998comments}; no new information can be gained. We note that this number  depends only on the relative object size, and does not depend on the object permittivity. We also note that this result has previously been derived in the framework of functional analysis by Bucci and Isernia \cite{bucci1997electromagnetic}.

\section{Discussion}\label{discussion}

\subsection{Implications on the Forward scattering problem}

The Nyquist-Shannon sampling theorem \cite{shannon1949communication} states that for a function that is bandlimited to a maximum frequency component, $f_b$, the minimum sampling frequency required to reconstruct the function is given by $f_s = 2f_b$. Applying this theorem to the scattered electric field, which is known to be bandlimited, implies that for a fixed object size, $a$, it is optimal to make $4k_b a$ equally spaced measurements of the scattered field. In an experimental setup it may be necessary to measure the  fields scattered due to an object. It is common to have a dedicated antenna for each measurement; we thus have a lower bound on the hardware complexity of the experiment. Further, if it is desired to estimate the scattered field at more points than these, a suitable interpolation scheme can be applied to the sampled values of the field \cite{candocia1998comments}.

A setup such as that described above---surrounding an object with several antennae---is fairly common in experiments which perform breast cancer detection using microwave imaging techniques \cite{klemm2009radar}. Although our results apply to a two-dimensional geometry, the same approach can be easily applied to the three-dimensional problem. 

\subsection{Implications on the Inverse problem}

In the inverse scattering problem, the measurements of the scattered  fields are typically noise corrupted. Also, it has been shown that the scattered field is bandlimited to $2k_ba$. Thus, a robust strategy for determining the unknown permittivity of an object would be to: (i) Take the DFT of the measurements (i.e.~obtain $\tilde{v}_k,\,k\in[0,M-1]$), (ii) apply a bandpass filter which zeroes out all $\tilde{v}_k$ components for $\lceil 2k_ba \rceil <k< M - \lceil 2k_ba \rceil$, and (iii) use the remaining $2 \lceil 2k_ba \rceil$ components of $\tilde{v}_k$ in Eq.~(\ref{dftscatf}) for the purpose of solving the inverse problem, i.e. determining the dielectric contrast, $\chi_n$. More constraints on $\chi_n$ via Eq.~\eqref{dftscatf} can be imposed by increasing the number of incidence angles, but as shown earlier (Section \ref{multiple}), the maximum number of useful incidence angles is limited to $ \lceil 2k_ba \rceil$ as well.

Thus, the property of bandlimitedness can be used to make the inverse problem noise-resilient to a certain degree.  
\appendices

\section{Bessel Functions of Fixed Argument} \label{Bessel}

\begin{figure}[htbp]
\begin{center}

\includegraphics[width=0.5\textwidth]{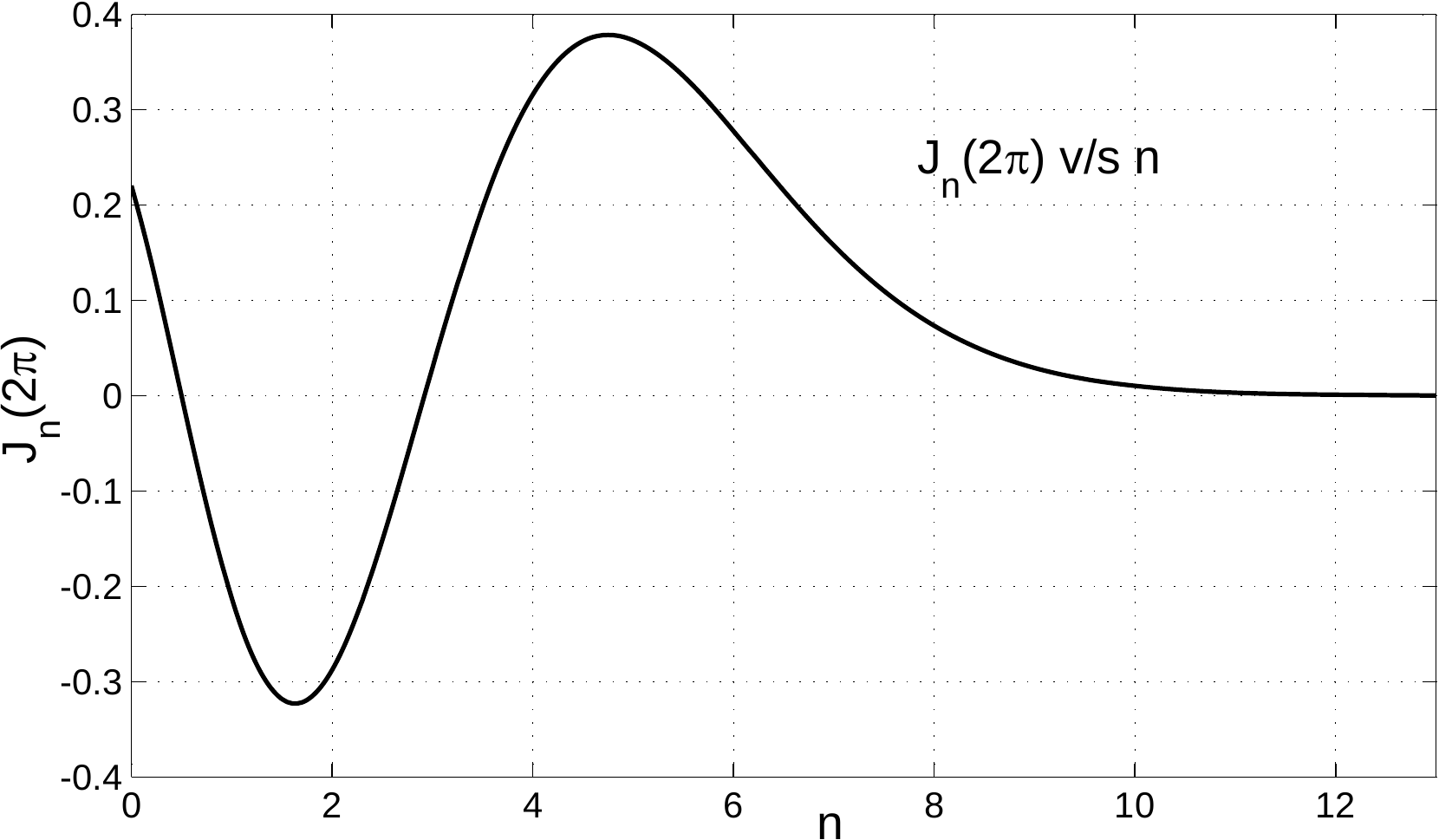}
\caption{Bessel function $J_n(x)$ as a function of $n$ for $x=2\pi$.}
\label{Jnx}

\end{center}
\end{figure}

In this section, we consider the properties of the $n^{\text{th}}$ order Bessel function for a fixed argument, $x$, i.e.~$J_n(x)$, where $n,x\in\mathbb{R}$. We are primarily interested in explaining a key feature of Figure \ref{Jnx}: the observation that $J_n(x)$ has a steep and monotonic decay in amplitude  beyond a certain threshold value of $n$ for a fixed $x$. This threshold value is subsequently identified.

\begin{figure}[htbp]
\begin{center}

\includegraphics[width=0.5\textwidth]{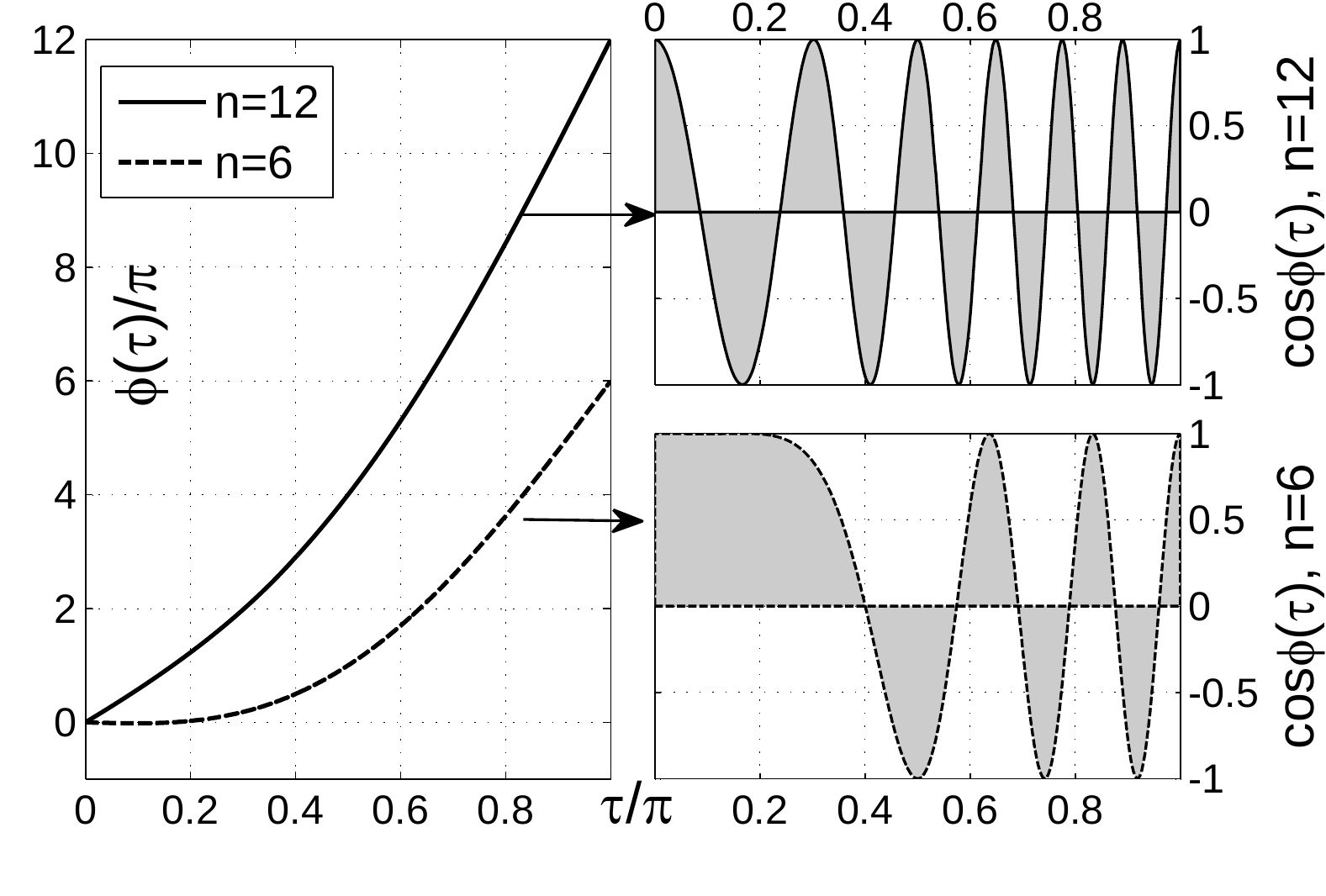}
\caption{Phase $\phi(\tau)=n\tau-x\sin\tau$ as a function of $\tau$ for $x=2\pi$ and two values of $n$: $n=12$ and $n=6$ (left), $\cos\phi(\tau)$ for $n=12$ (top right), and  $\cos\phi(\tau)$ for $n=6$ (bottom right).}
\label{phitau}

\end{center}
\end{figure}

We start by considering the integral representation of $J_n(x)$ for integer $n$ as
\begin{equation}
\label{BesselIntegral}
J_n(x) = \frac{1}{\pi}\int_0^\pi \cos(n\tau-x\sin\tau)\,d\tau.
\end{equation}

The argument of the cosine term in the above expression, $\phi(\tau)=n\tau-x\sin\tau$ has been plotted in Figure  \ref{phitau} for two values of $n$ at $x=2\pi$; $n=6$ and $n=12$. It is clear from Figure \ref{Jnx} that while $J_6(2\pi)$ is far from being zero, $J_{12}(2\pi)\approx 0$. This is graphically illustrated in Figure \ref{phitau}, where $\cos\phi(\tau)$ is plotted for both values of $n$. It is evident that the area under the curve is clearly non-zero in the case of $n=6$ (bottom right of the figure), where as the positive and negative areas seem much better balanced in the case of $n=12$ (top right of the figure).

The above argument can be presented in a more abstract manner, as follows. For a fixed $x$, and as $n$ increases, the value of $|n/x|$ exceeds $1$ and grows. The phase function, $\phi(\tau)=n(\tau-\frac{x}{n}\sin\tau)$ has a linear component and a sinusoidal component, and as $|n/x|$ grows, the linear component dominates. As a result, the intervals in $\tau$ (denoted by $\Delta\tau_1$) where $\phi(\tau)$ goes from $m\pi$ to $m\pi+\frac{\pi}{2}$, begin to approach the length of the intervals in $\tau$ (denoted by $\Delta\tau_2$) where $\phi(\tau)$ goes from $m\pi+\frac{\pi}{2}$ to $(m+1)\pi$. The area under the $\cos\phi(\tau)$ curve for the above two intervals, thus become approximately equal while being opposite in sign, i.e.~for $m\in[0,n-1]$ :
\begin{equation}
\int_{\Delta\tau_1}\cos\phi(\tau)\,d\tau \approx - \int_{\Delta\tau_2}\cos\phi(\tau)\,d\tau.
\end{equation}
Thus by breaking up the total integral of Eq.~(\ref{BesselIntegral}) into such intervals, we arrive at the result that as $|n/x|$ grows beyond $1$, $J_n(x)$ monotonically approaches zero. By noting that this behaviour depends simply upon $|n/x|$, and by inspection of Figure \ref{phitau} it can be said that $J_n(x)\approx 0$ for $|n| \ge \lceil{2|x|}\rceil$. The modulus on $n$ come from the observation that $|J_n(x)|=|J_{-n}(x)|$.

\section{Discrete Fourier transform of a plane wave}
\label{planewaveDFT}
Consider a $z$-polarized plane wave propagating in the $x-y$ plane, measured on a circular contour in the plane. Let there be $M$ observation points on this circle of radius $r_0$, giving an observed electric field vector, $x$, as 
\begin{equation}
\label{planewave}
x[m] = \exp\left[-jk_0r_0\cos\left(\frac{2\pi m}{M}\right)\right],\, m\in[0,M-1],
\end{equation} 
for a plane wave travelling along the $+x$ axis. The discrete Fourier transform (DFT) of this vector is given by $\tilde{x}[k] =  \overset{M-1}{\underset{m=0}{\sum}}\, \exp[-j2\pi km/M] x[m]$, for $k\in[0,M-1]$, whose properties are now considered.

\subsection{Symmetry of the DFT}
\label{symmdft}
\label{planewaveDFT-symmetry}

We can simplify it's symmetric counterpart, $\tilde{x}[M-k]$ by noting that for the given $x$, we have $x[m]=x[M-m]$, which leads to: $\tilde{x}[M-k] =\overset{M-1}{\underset{m=0}{\sum}}\, \exp[j2\pi km/M] x[M-m]$.
Substituting $M-m=\mu$, we get 
\begin{eqnarray}
&{}&\tilde{x}[M-k] = \overset{1}{\underset{\mu=M}{\sum}}\, \exp\left[\frac{j2\pi k(M-\mu)}{M}\right] x[\mu] \times \nonumber \\
&{}& \overset{1}{\underset{\mu=M}{\sum}}\, \exp\left[\frac{-j2\pi k\mu}{M}\right] x[\mu] = \tilde{x}[k] 
\end{eqnarray}
The last equality follows because the summand at $\mu=M$ is the the same as the value for $\mu=0$ since the field $x$ is periodic with period $M$.

\subsection{Band Limited Nature}
To arrive at the band limited nature of the incident field, we use the Jacobi-Anger expansion, which states that
\begin{equation}
\label{JA}
\exp(jz\cos\theta)=\overset{\infty}{\underset{p=-\infty}{\sum}}\,j^pJ_p(z)\exp(jp\theta),
\end{equation}
and the following expression for the discrete delta function, 
\begin{equation}
\label{delta}
\delta[k-k^\prime] = \frac{1}{M}\overset{M-1}{\underset{m=0}{\sum}}\,\exp\left[-\frac{j2\pi mk}{M}\right]\exp\left[\frac{j2\pi mk^\prime}{M}\right],
\end{equation}
where $\{k,k^\prime\}\in[0,M-1]$. Observe that the transformation $k^\prime\rightarrow k^\prime+qM$, where $q\in\mathbb{Z}$, leaves the above equation unchanged.

The $k^{\text{th}}$ Fourier component of $x$ from Eq.~(\ref{planewave}), after applying Eqs.~(\ref{JA}) and (\ref{delta}), is 
\begin{eqnarray}\nonumber
\tilde{x}[k] &= &\overset{M-1}{\underset{m=0}{\sum}}
 \exp\left[\frac{-j2\pi mk}{M}\right] \\ \nonumber
 &{}& \times \overset{\infty}{\underset{p=-\infty}{\sum}}\,j^pJ_p(-k_0r_0) \exp\left[\frac{j2\pi mp}{M}\right] \\ 
 &=& \overset{\infty}{\underset{p=-\infty}{\sum}}\,j^pJ_p(-k_0r_0)\,M \delta[k-(p+qM)] \nonumber \\
& =  &\overset{\infty}{\underset{q=-\infty}{\sum}}\,j^{k-qM} M J_{k-qM}(-k_0r_0)
\end{eqnarray}
In other words, the $k^{\text{th}}$ Fourier component of $x$ comprises the $k^{\text{th}}$ order Bessel function and it's $M-$shifted orders, all of the same (fixed) argument. 

From the analysis of the properties of $J_k(-k_0r_0)$ as a function of $k$ in Appendix \ref{Bessel}, it is clear that the DFT of the incident plane wave is band-limited. This is  because for $k>\lceil{2k_0r_0}\rceil$, $J_k(-k_0r_0) \sim 0$, and therefore $\tilde{x}[k] \sim 0$. Thus the incident field on a contour of radius $r_0$ can be represented by $\lceil{2k_0r_0}\rceil$ coefficients in the DFT basis.

\end{document}